# Surface-wave photonic quasicrystal

Zhen Gao[1], Fei Gao[1], Youming Zhang[1], Hongyi Xu[1], Baile Zhang*[1,2]

*[1]Division of Physics and Applied Physics, School of Physical and Mathematical Sciences, Nanyang Technological University, Singapore 637371, Singapore*

*[2]Centre for Disruptive Photonic Technologies, Nanyang Technological University, Singapore 637371, Singapore*

*E-mail: blzhang@ntu.edu.sg (B. Zhang)*

## Abstract

In developing strategies of manipulating surface electromagnetic waves, it has been recently recognized that a complete forbidden band gap can exist in a periodic surface-wave photonic crystal, which has subsequently produced various surface-wave photonic devices. However, it is not obvious whether such a concept can be extended to a non-periodic surface-wave system that lacks translational symmetry. Here we experimentally demonstrate that a surface-wave photonic quasicrystal that lacks periodicity can also exhibit a forbidden band gap for surface electromagnetic waves. The lower cutoff of this forbidden band gap is mainly determined by the maximum separation between nearest neighboring pillars. Point defects within this band gap show distinct properties compared to a periodic photonic crystal for the absence of translational symmetry. A line-defect waveguide, which is crafted out of this surface-wave photonic quasicrystal by shortening a random row of metallic rods, is also demonstrated to guide and bend surface waves around sharp corners along an irregular waveguiding path.

Photonic crystals (PCs), periodic systems with forbidden band gaps for electromagnetic (EM) waves in all directions [1], have attracted great interests in the past decades due to their appealing scientific and engineering applications. However, because PCs are based on Bragg interferences, PC devices are generally limited to the wavelength scale. On the other hand, surface EM waves at the metal/dielectric interfaces ["surface plasmons" (SPs) at optical frequencies] [2-3] offer the potential for controlling light on a subwavelength scale because of their intrinsic subwavelength nature. Although SPs in the visible/near-infrared spectra suffer from severe metallic loss, spoof SPs [4-15], or the surface EM waves at frequencies ranging from microwave, terahertz to far-infrared, hold promising application prospects because of their outstanding waveguiding performance with negligible loss. Recently, by merging the subwavelength feature of spoof SPs with the forbidden band gap of PCs, some unconventional surface-wave functionalities and devices based on surface-wave PCs [16-18] are developed, such as multi-directional splitting of surface waves with full bandwidth isolation [19], efficient broadband guiding and bending of surface waves along a line-defect waveguide [20], and ultra-slow-wave devices [21].

While all the previous surface-wave devices utilizing the photonic band gap were demonstrated in periodic surface-wave PCs [16-21], whether a similar band gap can exist in a non-periodic surface-wave system still remains unknown. Surface-wave PCs are unique in the sense that they combine the properties of conventional periodic PCs [1] and locally resonant wire-medium metamaterials [17]. In a conventional PC, the interaction between a unit cell with its surrounding periodic unit cells (e.g. the interaction between a rod and its surrounding rods) forms Bragg scattering, which is the very origin of the band gap. However, in a locally resonant wire-medium metamaterial [17], the resonance frequency is fully determined by a single wire itself, with almost zero influence from the surrounding wires. In this case, the wires can be even randomly placed without affecting the resonance frequencies. Being different from a conventional PC and a locally resonant wire-medium metamaterial, in a surface-wave PC, a unit cell exhibits both local resonance and near-field interaction with its neighboring unit cells. It is thus worthwhile to verify if the periodicity is the necessary condition of a band gap for manipulating surface wave propagation.

Inspired by a photonic quasicrystal [22-26] whose crystal lattice lacks the translational symmetry, here we experimentally demonstrate that a surface-wave photonic quasicrystal with Penrose tiling [27] also possesses a complete band gap for surface EM waves. The absence of the translational symmetry in Penrose lattice allows different resonance frequencies and localization properties of defect cavity resonance modes at different defect positions. Moreover, efficient guiding and bending of surface waves along a line-defect waveguide by shortening a random row of metallic rods in this surface-wave photonic quasicrystal are also demonstrated with transmission spectra measured and mode profiles imaged directly.

We first consider an array of circular aluminum rods with radius $r = 1.25$ mm and height $h = 5$ mm on a flat aluminum plate, as shown in Fig. 1(a), where each rod locates at a vertex of the two-dimensional (2D) Penrose lattice. The edge of each rhomb is $d = 5$ mm. The Penrose lattice utilizes two different types of tiles to fill the whole 2D plane: a thin rhomb with vertex angles of 36º and 144º, and a fat rhomb with vertex angles of 72º and 108º, as shown in the inset of Fig. 1(a).

To experimentally demonstrate the existence of a band gap in the surface-wave photonic quasicrystal, we measured the transmission spectrum of surface waves supported on this quasicrystal structure at microwave frequencies. Two homemade monopole antennas (with dimensions of 1.5 mm in diameter and 5 mm in length), connected to a vector network analyzer (VNA) R&S ZVL-13, were used as the source and the probe placed at opposite sides of the surface-wave photonic quasicrystal to measure the transmission coefficient (S-parameters S21) of surface waves. Note that because of the impedance mismatch, the monopole antennas are inefficient and mainly sensitive to the evanescent waves. The measured transmission coefficient should be understood as a relative transmission measurement. Fig. 1(b) shows the measured transmission coefficient between the source and probe, from which we can clearly observe a forbidden band gap starting from 12.6 GHz (the upper edge of the band gap is not shown because of the frequency range limit of VNA), within which the relative transmission from -40 dB to -35 dB is at the noise level. For comparison, we then measured the transmission coefficient of a previous periodic surface-wave PC [18-20], as shown in Fig. 1(c), with periodicity $d = 5$ mm. The measured transmission spectrum of the periodic surface-wave PC is presented in Fig. 1(d), which also exhibits a wide forbidden band gap starting from 12.6 GHz. Evidently, almost the same photonic band gap exists for both periodic surface-wave PCs and non-periodic surface-wave photonic quasicrystals, indicating that periodicity is not a necessary condition of the forbidden band gap for surface EM waves.

To construct a defect cavity in the surface-wave photonic quasicrystal, we replace one metallic rod with a slightly shorter one [18]. Due to the absence of translational symmetry, the properties (resonance frequency, localization and field pattern) of the defect cavity can be very different from those in a periodic surface-wave PC [18]. We first slightly shorten a metallic rod from $h = 5$ mm to $h_d = 4.3$ mm at the center of the surface-wave photonic quasicrystal [labeled as A in Fig. 2(a)]. This defect cavity is symmetrically surrounded by five nearest neighboring metallic rods. With the source placed in the defect cavity and the probe placed above the defect cavity, the measured near-field response spectrum of this surface defect cavity [red line in Fig. 2(b)] reveals a resonance mode at frequency $f_A = 13.75$ GHz located within the forbidden band gap of the surrounding surface-wave photonic quasicrystal (purple region). For comparison, by maintaining the positions of the source and probe, we also measured the near-field response spectrum of the perfect surface-wave photonic quasicrystal without any defect [black line in Fig. 2(b)]. No resonance peak can be observed. We also imaged the near-field profile ($E_z$) of the surface defect resonance mode using a microwave near-field scanning system, as shown in Fig. 2(c). The five-fold rotational symmetry is clearly observed. This further confirms that a surface defect cavity mode is created at the defect site.

If we shorten a metallic rod at another position [labeled as B in Fig. 2(a)] from $h = 5$ mm to $h_d = 4.3$ mm, the measured near-field response spectrum [green line in Fig. 2(b)] and near-field distribution [Fig. 2(d)] are quite different from those of the defect state at site A due to the change of the surrounding environment of the shortened rod. We now shorten a metallic rod at the third position [labeled as C in Fig. 2(a)] from $h = 5$ mm to $h_d = 4.3$ mm. This results in another different defect state whose measured near-field response spectrum and near-field distribution are presented in Fig. 2(b) (blue line) and Fig. 2(e), respectively. We note that shortening the metallic rod at site A corresponds to a defect cavity state with lower resonance frequency ($f_A = 13.75$ GHz) while shortening metallic rods at sites B and C corresponds to higher resonance frequencies ($f_B = 14.21$ GHz and $f_C = 14.52$ GHz). This is because at site A the five nearest neighboring rods are placed around the defect rod with equal distance. At site B, however, one nearest neighboring rod has been closer to the defect rod than other surrounding rods, producing a higher resonance frequency. At site C, two nearest neighboring rods are closer to the defect rod, giving rise to an even higher resonance frequency.

Using the same experimental setup, we now move on to construct more complex functionalities in surface-wave photonic quasicrystals: a quasi-straight line-defect waveguide and a line-defect waveguide with a sharp corner. We first construct a quasi-straight line-defect waveguide by shorting a row of aluminum rods from $h = 5$ mm to $h_d = 4.3$ mm, as shown in the inset of Fig. 3(a). Then we measured the relative transmission coefficient of the waveguide, as shown in Fig. 3(a) (red line). Compared with the transmission spectrum of a perfect surface-wave photonic quasicrystal without line-defect [black line in Fig.3 (a)], we can clearly observe a waveguiding band within the band gap of the surrounding surface-wave photonic quasicrystal (purple region) starting from 12.6 GHz. Measured spatial distribution of $E_z$ component in a transverse *xy*-plane 2 mm above the waveguide is presented in Fig. 3(b), which shows that surface waves efficiently propagate along the waveguide. For comparison, by removing all surrounding un-shortened rods, we also measured the transmission spectrum through only the defect rods (blue line) which works as a quasi-straight domino plasmon waveguide [8]. Because of the absence of surrounding rods, the surface-wave propagation is not affected by the band gap boundary at 12.6 GHz. The cutoff at around 14 GHz is an intrinsic polaritonic property of the domino plasmons.

We then demonstrate a line-defect waveguide with a sharp corner. As shown in the inset of Fig. 4(a), a sharp corner in the surface-wave photonic quasicrystal can be created by shortening a row of aluminum rods in *x* and *y* directions successively. The measured transmission spectrum of the sharp corner is shown in Fig. 4(a) (pink line), a transmission band is observed within the forbidden band gap of the surrounding surface-wave photonic quasicrystal, being similar to that of the quasi-straight line-defect waveguide presented in Fig. 3(a). For comparison, the transmission through the perfect surface-wave photonic quasicrystal is also plotted in Fig. 4(a) (black line), which shows low transmission values. As presented in Fig. 4(b), the measured electric field distribution in a transverse $xy$-plane 2 mm above the sharp corner shows clearly that surface waves can be efficiently transmitted through a sharp corner in a surface-wave photonic quasicrystal, being similar to the wave bending phenomena in a periodic surface-wave PC [20]. In contrast, when all surrounding rods are removed, the transmission through the sharp corner greatly drops, as presented in Fig. 4(a) (blue line).

For completeness, we discuss the situation of a completely random structure with numerical simulation. It has been reported that both the density and spatial order have no influence on the forbidden band gap of a locally resonant wire-medium metamaterial [17]. Even in a random lattice [28] the forbidden band gap is kept the same with that of a periodic lattice. However, the situation is different in surface-wave PCs. We first simulate the influence of the change of periodicity on the forbidden band gap of a periodic surface-wave PC, as shown in Fig. 5(a), where five different values of periodicity $d$ are considered. We can observe that the cutoff frequencies (indicated with dashed lines) of the band gap of periodic surface-wave PCs are red shifted with increasing the periodicity from $d = 3$ mm (black line) to $d = 5$ mm (red line). This indicates that the density of surface-wave PC influence its forbidden band gap, which is in contrast to the hybrid locally resonant wire-medium metamaterials [17,28] whose forbidden band gap is determined only by the resonant nature of their constitutive unit cell (length of the metallic wire) rather than the density or periodicity. That is because each pillar in surface-wave PCs not only exhibits local resonance but also near-field interaction with its neighboring pillars.

After we have shown that the cutoff frequencies of the forbidden bang gaps of periodic surface-wave PCs are influenced by their densities, we simulate the transmission spectra of non-periodic surface-wave PCs with Penrose lattice and a completely random lattice to find what determines the cutoff frequencies of their forbidden band gaps. The completely random lattice, as shown in the inset of Fig. 5(b), has the same size and the same number of rods with the periodic surface-wave PC in Fig. 1(c).

As shown in Fig. 5(b), we observe that the simulated cutoff frequencies of periodic lattice (red dashed line) and Penrose lattice (green dashed line) almost overlap with each other, matching well with experimental results presented in Fig. 1. On the other hand, the cutoff frequency of the random lattice (blue dashed line) is red shifted compared to those of periodic and Penrose lattice. This is because, the maximum separation between nearest neighboring metallic pillars in the periodic surface-wave PC and the surface-wave photonic quasicrystal with Penrose lattice is the same ($d = 5$ mm). However, in the completely random lattice, the maximum separation between nearest neighboring metallic pillars, for a local metallic pillar, is larger than $d$, thus inducing lower cutoff frequency. We can thus conclude that the cutoff frequencies of both the periodic and non-periodic surface-wave PCs are determined by the maximum separation between nearest neighboring metallic pillars.

In conclusion, we have experimentally demonstrated that a photonic band gap can exist in a surface-wave photonic quasicrystal that composed of the Penrose tiling of aluminum pillars on a flat metal surface. This shows that periodicity is not a necessary condition for the existence of surface-wave photonic band gap. The cutoff frequencies of both the periodic and non-periodic surface-wave PCs are determined by the maximum separation between nearest neighboring metallic pillars. Moreover, efficient guiding and bending of surface waves in surface-wave photonic quasicrystals are also demonstrated.

**Acknowledgements**

This work was sponsored by the NTU Start-Up Grants, Singapore Ministry of Education under Grant No. MOE2015-T2-1-070 and MOE2011-T3-1-005.

## Figures and captions

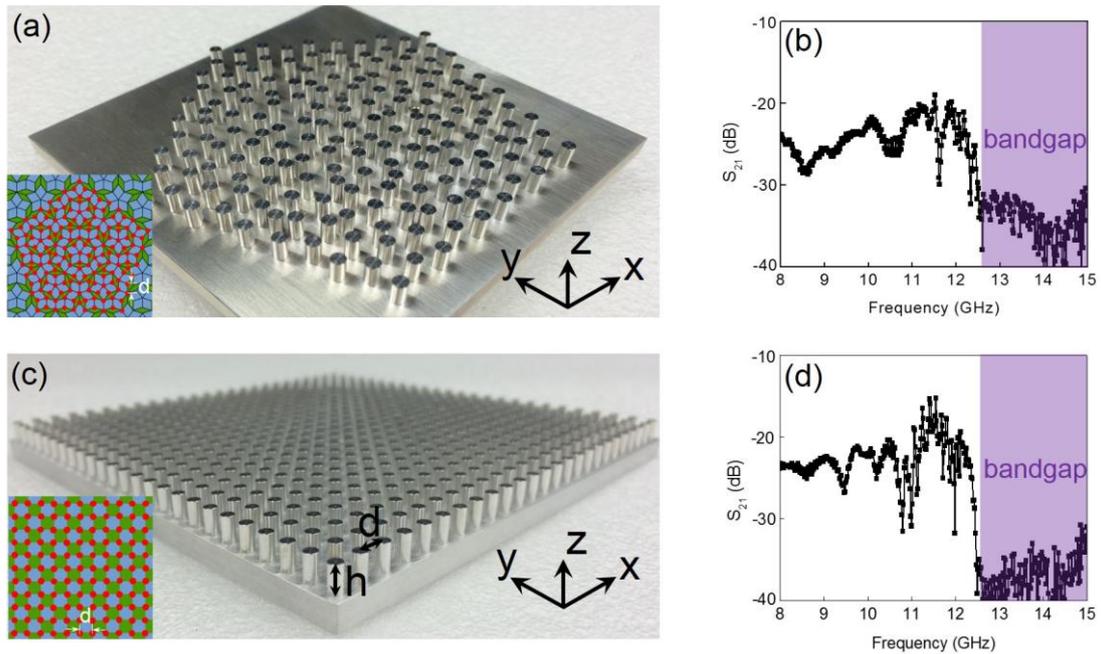

**FIG. 1.** (a) Photo of a 2D Penrose-type surface-wave photonic quasicrystal with aluminum rods (radius $r = 1.25$ mm, height $h = 5$ mm) placed at the vertices of thin and fat rhomb titles on a flat aluminum plate. The length of the rhomb side is $d = 5$ mm. (b) Measured transmission spectrum (S-parameter S21) of the surface-wave photonic quasicrystal. (c) Photo of the traditional periodic surface-wave PCs that consists of a square array of circular aluminum rods with radius $r = 1.25$ mm, height $h = 5$ mm, and periodicity $d = 5$ mm. (c) Measured transmission spectrum (S-parameter S21) of the periodic surface-wave PCs.

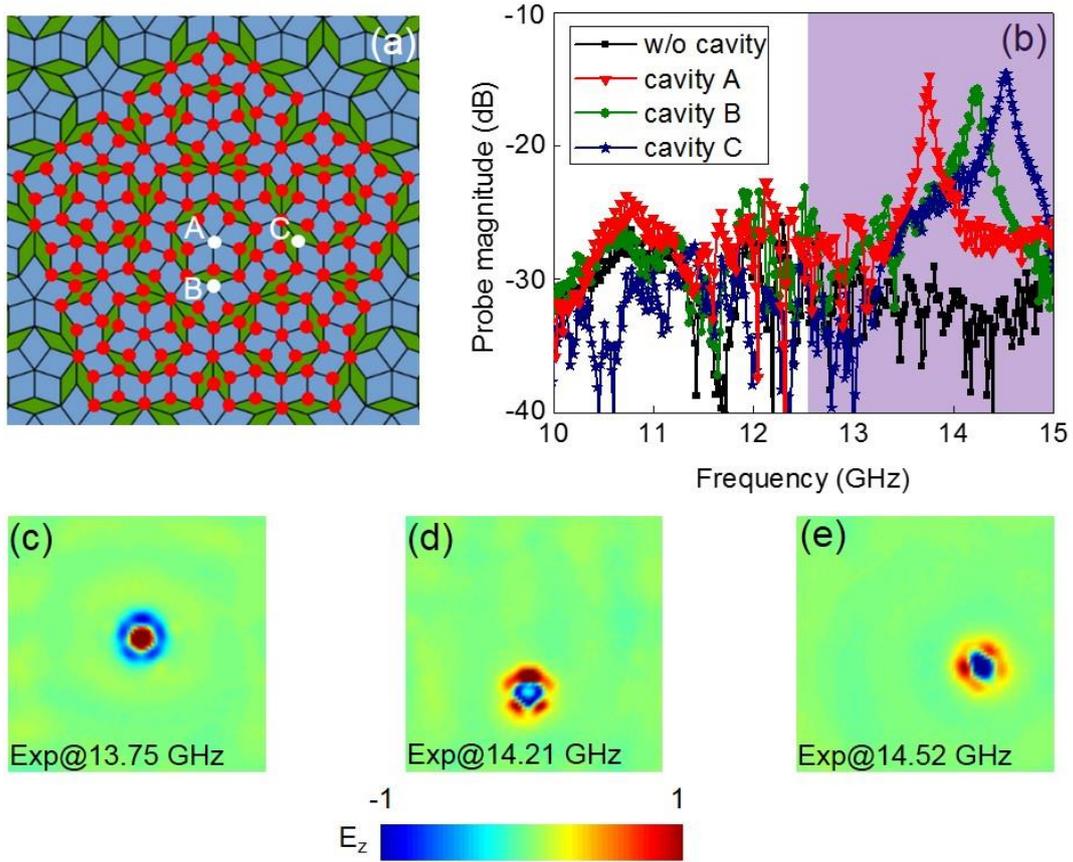

**FIG. 2.** (a) Schematic figure showing the quasicrystal (Penrose lattice) arrangement of aluminum rods (red dots) on a flat aluminum surface. Three defects (white dots) labeled as A, B, C are induced by shortening a metallic rod from $h = 5$ mm to $h_d = 4.3$ mm at each site. (b) Measured near-field response spectra of a surface-wave photonic quasicrystal with a single point defect (A: red line, B: green line, C: blue line) and without any defect (black line). (c) Observed field pattern ($E_z$) of defect cavity A at the resonance frequency 13.75 GHz. (d) Observed field pattern ($E_z$) of defect cavity B at the resonance frequency 14.21 GHz. (e) Observed field pattern ($E_z$) of defect cavity C at the resonance frequency 14.52 GHz.

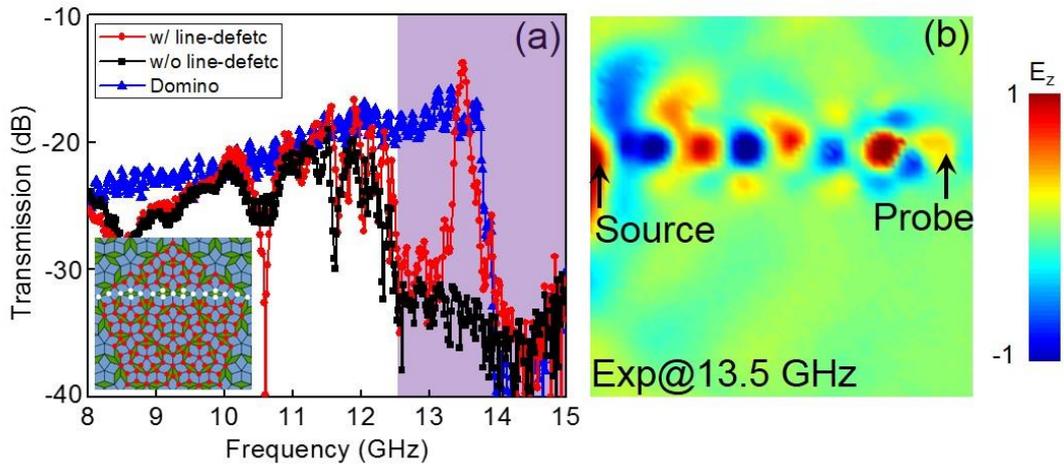

**FIG. 3.** (a) Measured transmission spectra of a quasi-straight line-defect waveguide (red line) and a perfect surface-wave photonic quasicrystal without line-defect (black line) as well as a quasi-straight domino plasmon waveguide (blue line) after removing surrounding un-shortened pillars. Inset shows the line-defect waveguide configuration. (b) Measured field pattern ($E_z$) of the quasi-straight line-defect waveguide at 13.5 GHz. Two monopole antennas as the source and probe are indicated with a pair of black arrows.

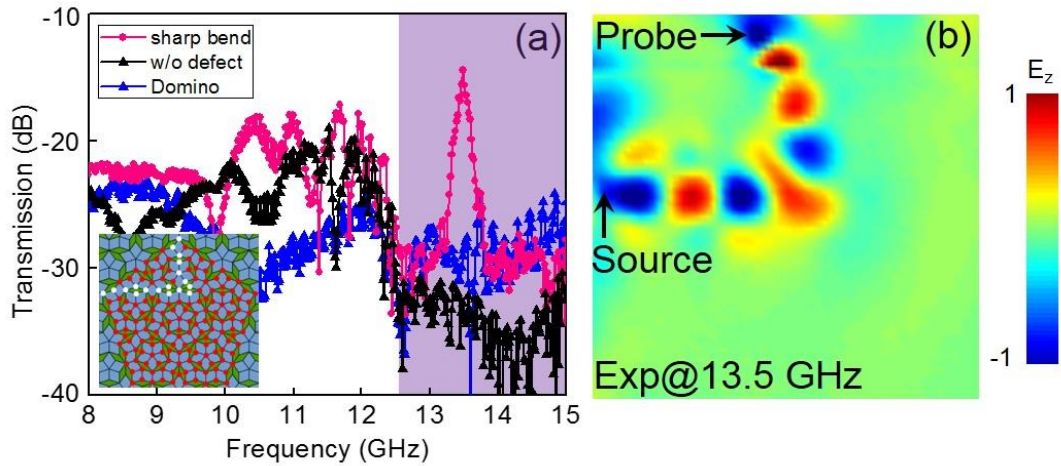

**FIG. 4.** (a) Measured transmission spectra of a bent line-defect waveguide (pink line) and a perfect surface-wave photonic quasicrystal without any defect (black line), as well as a bent domino plasmon waveguide with a sharp corner (blue line) after removing surrounding un-shortened pillars. Inset shows the line-defect waveguide configuration. (b) Observed field pattern ($E_z$) above the sharp corner at 13.5 GHz. Two monopole antennas as the source and probe are indicated with a pair of black arrows.

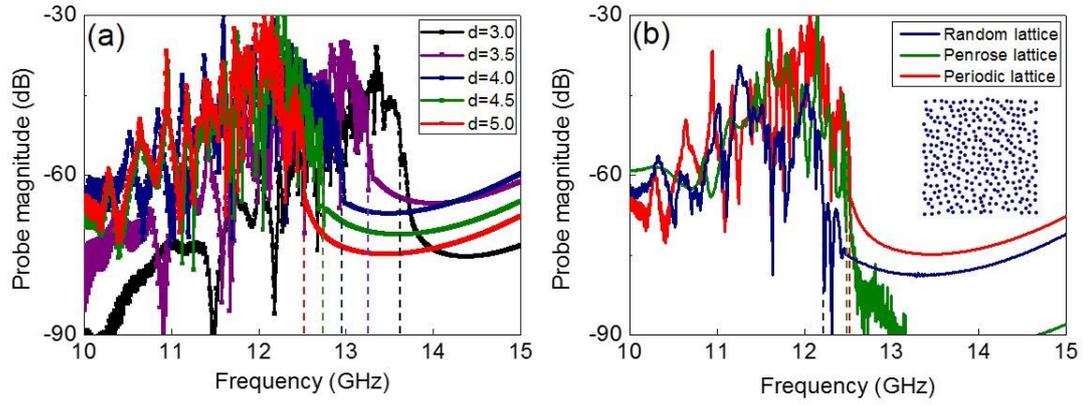

**FIG. 5.** (a) Simulated transmission spectra of periodic surface-wave PCs with different periodicity *d* equal to 3.0 mm (black line), 3.5 mm (purple line), 4.0 mm (blue line), 4.5 mm (green line), and 5.0 mm (red line). (b) Simulated transmission spectra of a periodic surface-wave PC (red line), a surface-wave photonic quasicrystal with Penrose lattice (green line) and a surface-wave random lattice (blue line). Inset shows the schematic of the random lattice.